\documentclass[twocolumn,showpacs,preprintnumbers,amsmath,amssymb]{revtex4}

\usepackage{epsfig}
\usepackage{dcolumn}
\usepackage{bm}

\begin{document}

\title{Splitting electrons into quasiparticles with a fractional edge-state
Mach-Zehnder interferometer}

\author{V.V. Ponomarenko}
\affiliation{Center of Physics, University of Minho, Campus
Gualtar, 4710-057 Braga, Portugal}
\author{D.V. Averin}
\affiliation{Department of Physics and Astronomy,
University of Stony Brook, SUNY, Stony Brook, NY 11794}

\date{\today}


\begin{abstract}
We have studied theoretically the tunneling between two
edges of quantum Hall liquids (QHL) of different filling
factors, $\nu_{0,1}=1/(2 m_{0,1}+1)$, with $m_0 \geq
m_1\geq 0$, through two separate point contacts in the
geometry of Mach-Zehnder interferometer \cite{mz1}. The
quasi-particle formulation of the interferometer model
is derived as a dual to the initial electron model, in
the limit of strong electron tunneling reached at large
voltages or temperatures. For $m\equiv 1+m_{0}+m_{1}>1$,
the tunneling of quasiparticles of fractional charge
$e/m$ leads to non-trivial $m$-state dynamics of
effective flux through the interferometer, which
restores the regular ``electron'' periodicity of the
current in flux despite the fractional charge and
statistics of quasiparticles. The exact solution
available for equal times of propagation between the
contacts along the two edges demonstrates that the
interference pattern of modulation of the tunneling
current by flux depends on voltage and temperature only
through a common amplitude.
\end{abstract}


\maketitle

\section{Introduction}

Electronic Mach-Zehnder interferometers (MZIs) based on
the integer quantum Hall states have been designed and
studied in recent experiments \cite{mz1,mz01}. This
device consists of two tunneling contacts between two
single-mode edges of the two-dimensional (2D) electron
liquid in the regime of the integer quantum Hall effect,
which are arranged to propagate effectively in the same
direction. The interferometer enables one to observe
pronounced interference patterns in the tunneling
current. In anticipation of possible realization of
similar interferometer based on the edges of electron
liquids in the regime of the fractional quantum Hall
effect (FQHE), MZI in this regime \cite{mz2,mz3} and
more complicated structures including it \cite{mz4} were
studied theoretically in search for signatures of the
fractional statistics of FQHE quasiparticles. Some of
these theories, however, (cf. Refs.~\cite{mz2} and
\cite{mz3}) were based on different postulated models of
the quasiparticle transport in MZI and obtained
conflicting result, e.g., different periods of the
tunnel current modulation by external magnetic flux
$\Phi_{ex}$ through the interferometer.

In this work, whose main results have been briefly
presented in Ref.~\cite{usprl}, we consider tunneling
between two edges of quantum Hall liquids (QHL) of in
general different filling factors, $\nu_{0,1}=1/(2
m_{0,1}+1)$ with $m_0 \geq m_1\geq 0$, through two
separate point contacts in the MZI geometry, and derive
its quasi-particle model from the electronic description
of the interferometer. The latter is always correct in
the limit of weak tunneling, when the two edges are well
separated from each other, and only whole electrons can
be transferred between them through opaque tunnel
barrier which itself does not contain FQHE liquid. Using
the scaling growth of electron tunneling amplitudes with
increasing voltage (or temperature), we demonstrate that
the quasi-particle formulation of the interferometer
model emerges naturally as a dual to the initial
electron model in the limit of strong electron
tunneling. This model shows that backscattering at the
two interferometer contacts, which is weak for strong
electron tunneling, produces quasiparticles of the same
charge $e_X=2\nu_0\nu_1/(\nu_0+\nu_1)$ as in the
situation of one point contact between the edges
\cite{b14}. If the filling factors of the two edges are
equal, the point-contact quasiparticles coincide
\cite{b15} with the "bulk" Laughlin quasiparticles, as
has been confirmed in the shot-noise experiments
\cite{noiseexper}.

The duality transformation used in this work to derive
quasiparticles in the MZI, and the resulting
quasiparticle tunnel Hamiltonian, are very similar to
those in our previous treatment \cite{b9} of the antidot
tunneling between fractional quantum Hall liquids
(FQHLs) with different filling factors. Both systems
exhibit an unusual phenomenon: Interference phase
accumulated between the two point contacts is not
determined solely by an external magnetic flux
$\Phi_{ex}$ confined between the two edges but has a
statistical contribution which transforms $\Phi_{ex}$
into an effective flux $\Phi$. In the interferometer,
each electron tunneling changes $\Phi$ by $\pm m\Phi_0$,
where $m=1+m_0+m_1$ and $\Phi_0=2\pi \hbar c/e$ is a
flux quantum equal to $2\pi$ in the units ($\hbar,
c,e=1$) we use in this paper, whereas in the case of the
antidot tunneling, the similar factor $m$ is given by
$m=m_0-m_1$. (This difference reflects the difference in
the edge propagation in the two structures. The two
edges propagate in the same direction in the MZI, and in
the opposite directions in the antidot.) As a result of
this flux change, the system acquires $m$ different
quantum states, whose effective fluxes $\Phi$ differ
from each other by $\Phi_0$ modulo $m\Phi_0$. These
states can not be coupled by perturbative electron
tunneling and therefore do not show up in the
weak-tunneling ``electron'' regime. In the
non-perturbative regime of strong tunneling, however,
the states become mixed as $\Phi$ is changed by one flux
quantum $\pm \Phi_0$ in the course of tunneling of
individual quasiparticles. The charge transfer
associated with this flux change, $e/m=e_X$, gives the
fractional charge of the quasiarticles which, in MZI,
coincides with the usual point-contact quasiparticles in
one point contact. In this respect, the MZI is different
from the antidot formed by FQHLs with different filling
factors, where the tunneling quasiparticles are
different from those in one point-contact, but can be
constructed from them through the process of multiple
interference \cite{b9}. Our derivation of the
quasiparticle Lagrangian in this work is a mathematical
demonstration of such a splitting of electron into
quasiparticles by the dynamics of flux. In the
particular case of coincident filling factors,
$\nu_0=\nu_1$, the model we derive agrees with the
quasi-particle model assumed in Ref.~\cite{mz3}. Our
result also confirms that the quasi-particle model of
Ref.~\cite{mz2} does not correspond to electron
tunneling at two separate point contacts in the
weak-tunneling limit, and probably does not represent
any realizable geometry of an interferometer.

In the situation of symmetric interferometer, when the
times $t_0$ and $t_1$ of propagation between the
contacts along the two edges are equal: $\Delta t\equiv
(t_0-t_1)/2 = 0$, the quasiparticle Lagrangian can be
solved by the methods of exactly solvable models. The
resultant expression for the tunneling current can also
be used for $V,T < 1/\Delta t$. This exact expression
describes the crossover from the regime of electron to
quasiparticle tunneling with increasing voltages or
temperatures. The tunneling conductance vanishes in both
of the two limits of large and small voltages and/or
temperatures. The large-voltage behavior of the exact
tunneling conductance agrees to the leading order in
large $V$ with the conductance found in Ref.~\cite{mz3},
limiting the validity of the quasiparticle calculation
in Ref.~\cite{mz3} to this order. The conductance
reaches its maximum of about $e^2/(2\pi \hbar m)$ in the
crossover region between the regimes of electron and
quasiparticle tunneling. The conductance peak extends
between the energies defined by the bigger and the
smaller of the two point contact tunneling amplitudes,
and therefore the peak width increases with increasing
asymmetry between the two amplitudes. This asymmetry
also makes the peak height larger, approaching more
closely the saturation value $e^2/(2\pi \hbar m)$. In
contrast to this, the magnitude of interference
conductance oscillations as a function of the magnetic
flux $\Phi_{ex}$ decreases steadily with increasing
ratio of the two tunneling amplitudes. The oscillations
should have the perfect 100\% visibility in the
interferometer with the identical point contacts.

\begin{figure}[htb]
\setlength{\unitlength}{1.0in}
\begin{picture}(3.,2.2)
\put(0.3,-0.1){\epsfxsize=2.3in \epsfbox{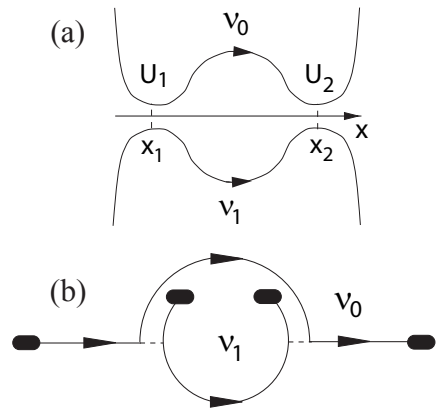}}
\end{picture}
\caption{Mach-Zehnder interferometer considered in this work. (a)
Conceptual diagram of the two co-propagating edges of QHLs with
different filling factors $\nu_0$ and $\nu_1$ coupled at points
$x_j$, $j=1,2$, by two point contacts with tunneling amplitudes
$U_j$. The edges are assumed to support one bosonic mode each, with
arrows indicating direction of propagation of these modes. (b)
Schematic geometry of the edge propagation in the experimentally
realized interferometers \protect  \cite{mz1,mz01}. Filled areas
denote the Ohmic contacts which emit/absorb edge modes, and special
arrangement of which makes it possible to implement tunneling between
the co-propagating edges within one plane of the two-dimensional
electron gas. }
\end{figure}

The paper is organized as follows. Section II defines
the electron tunneling model considered in this work and
presents perturbative calculation of the electron
tunneling current and integral visibility of its
interference pattern in the regime of weak electron
tunneling. Section III treats the electron tunneling
model in the opposite limit of strong coupling in both
contacts of the interferometer. We describe the
bosonization procedure for the Klein factors of electron
tunneling operators which implements the flux
attachment, and develop the instanton transformation
leading to the dual model of quasiparticle tunneling. In
the perturbative regime of weak quasiparticle tunneling,
$V\Delta t \gg 1$ or $T \Delta t \gg 1$, we calculate
the dc current. Section IV presents perturbative
calculations of the shot noise in both limits of weak
electron and weak quasiparticle tunnelings. In Section
V, we consider symmetric interferometer and obtain exact
solution of its quasiparticle model, through
fermionization for $m=2$ or by Bethe-ansatz technique
for general $m$. This solution is used to calculate the
average tunneling current in the interferometer and to
analyze its dependence on the magnetic flux, voltage,
and temperature.

\section{Electron tunneling model of the Mach-Zehnder
interferometer}

\subsection{Description of the edge states}

To formulate the effective electronic model of the MZI (Fig.~1) we
adopt the standard bosonization description of each of the two
single-mode edges with filling factors $\nu_l=1/ (2m_l+1)$, $l=0,1$.
In this description, the electron operator $\psi_l$ of the edge $l$
is expressed as \cite{b12}
\[ \psi_l=(D/2\pi v_l)^{1/2} \xi_l e^{i[\phi_l(x,t)/
\sqrt{\nu_l} +k_lx] }\,  . \]
Here $\phi_l$ are the two bosonic modes propagating in the same
direction (in Fig.~1, to the right) with velocities $v_l$ taken to
be positive, $v_l>0$, the Majorana fermions $\xi_l$ account for
mutual statistics of electrons in different edges, and $D$ is a
common large-energy cut-off of the edge modes. The Fermi momenta
$k_l$ define the average electron density in the edges, while the
operators of the density fluctuations are:
\[ \rho_l(x,\tau)= (\sqrt{\nu_l}/2 \pi) \partial_x \phi_l(x, \tau).
\]

The standard quadratic Lagrangian of the bosonic fields $\phi_l$
defines their real-time correlators, which at finite temperature $T$
can be written as
\begin{equation}
\langle  \phi_l(x,t) \phi_l(0,0)  \rangle = -\ln\{
\delta \sinh(\pi T (x/v_l-t+i/D))\} \, , \label{e0}
\end{equation}
where $\delta$ comes from an infrared cut-off and should
be taken to zero at the end of calculations.
Substituting this expression into the standard
definitions of the retarded and advanced Green functions
$g^{R,A}(x,t)$ of these modes one finds:
\begin{eqnarray}
g^{R,A}(x,t) & = &  \mp i\theta(\pm t) \langle [\phi_l(x,t), \phi_l]
\rangle  \nonumber \\ & = & \pm \pi \theta(\pm t) \mbox{sgn}(x-v_l
t) \, .
\end{eqnarray}
The Fourier-transformed functions
\[ g^{R,A}(x,\omega)=\int dt e^{i\omega t} g^{R,A}(x,t) \]
satisfy the condition $g^{A}(x,\omega)=[g^{R}(-x,\omega)]^*$, and
are equal to
\[ g^{R,A}(x,\omega)=\frac{2 \pi}{i(\omega \pm i0) } \left(-
\frac{\mbox{sgn} (x)}{2} \pm \theta(\pm x)e^{i\omega x/v} \right) \,
. \]
Analytical continuation of these expressions according
to the standard prescription: $g(x,\omega)=
-g^R(x,\omega)|_{\omega \rightarrow i \omega}$ for
positive frequencies $\omega$, and $g(x,\omega)=
-g^A(x,\omega)|_{\omega \rightarrow i \omega}$ for
negative $\omega$, gives the Fourier transform of the
imaginary-time-ordered correlators (see, e.g.,
Ref.~\cite{b8})
\[ \int_0^{1/T} d\tau e^{i\omega \tau} \langle T_\tau\{\phi_l (x,\tau)
\phi_p(0,0) \}\rangle= \delta_{lp} g(x/v_l,\omega) \]
as follows:
\begin{equation}
g(z,\omega)=\frac{2 \pi}{\omega} \mbox{sgn} (z) \Big(-\frac{1}{2}
+\theta(\omega z) e^{-\omega z} \Big) \, . \label{e1}
\end{equation}
The first term on the right-hand-side of Eq.~(\ref{e1}) defines the
usual equal-time commutation relations
\[ [\phi_l(x),\phi_p(0)]=i \pi \mbox{sgn}(x)\, \delta_{lp} . \]

\subsection{Weak electron tunneling model of MZI}

With the bosonized electron operators, Langrangian
describing electron tunneling in the two contacts is:
\begin{equation}
{\cal L}_t = \sum_{j=1,2} [\frac{DU_j}{2\pi} e^{i\kappa_j} e^{i
\lambda \varphi_j} + h.c.] \equiv \sum_{j=1,2} (T_j^+ +T_j^-)\, ,
\label{e2}
\end{equation}
where $U_j$ and $\kappa_j$ are the absolute values and the phases of
the dimensionless tunneling amplitudes, and
\begin{eqnarray}
\lambda \varphi_j(t) \equiv \frac{\phi_0(x_j,t)}{\sqrt{\nu_0}} -
\frac{\phi_1(x_j,t)}{\sqrt{\nu_1} } \, , \nonumber  \\
\lambda=\left[ \frac{\nu_0 +\nu_1}{\nu_0 \nu_1 }\right]^{1/2}
=\sqrt{2m} \, . \label{aftere2}
\end{eqnarray}
The factor $\lambda$ is chosen in such a way that the
normalization of the bosonic operators $\varphi_j$
coincides with that of the fields $\phi_l$, so that the
imaginary-time correlators of $\varphi_j$ are given by
the same Eq.~(\ref{e1}) with $z=0$:
$g(0,\omega)=\pi/|\omega|$. The products of the Majorana
fermions $\xi_1 \xi_2$ were omitted from the Lagrangian
(\ref{e2}), since they cancel each other in each
perturbative order due to charge conservation. The
phases $\kappa_j$ include contributions from the
external magnetic flux $\Phi_{ex}$ through the
interferometer and from the average numbers $N_0$ and
$N_1$  of electrons accumulated, respectively, on the
two sides of the interferometer between its tunnel
contacts, so that
\[ \kappa \equiv \kappa_2 - \kappa_1=2\pi[(\Phi_{ex}/\Phi_0)+
(N_0/\nu_0)-(N_1/\nu_1)]+ \mbox{const}. \]
In practical devices \cite{mz1,mz01}, the external magnetic flux
$\Phi_{ex}$ is defined by the area enclosed between the propagating
edges, including the area of the 2D electron gas of one of FQHLs
(see Fig.~1b), which can be modified by a modulation gate. Note that
non-trivial arrangement of the edges and tunneling contacts in
practical devices shown in Fig.~1b is dictated by the confinement of
the MZI structure to the plane of one 2D electron gas. In principle,
more direct implementations of the electronic MZIs should be
possible in the double-layer structures.

When a bias voltage $V$ is applied to the interferometer, it creates
a difference between the electrochemical potentials of the edges and
also changes their local densities and hence the Fermi momenta. In
the bosonic-field Lagrangian, the first effect can be accounted for
by adding the time-dependent phase factors to the tunneling
operators, $T_j(t)^\pm \rightarrow T^\pm_j(t) \exp\{\mp iVt\}$ in
Eq.~(\ref{e2}), while the second effect should change the phases
$\kappa_j$. This means that the phase difference $\kappa$ is also a
function of the applied voltage $V$: $\kappa= \kappa(V)$. The
voltage-induced contribution to $\kappa$ depends on the
electrostatics of the interferometer, and on the way the voltage is
applied. For instance, if the voltage changes the electrochemical
potential of the edge $0$ only, and the charge density is not fixed
by electrostatics due to effective screening by an external gate,
the phase varies as $\kappa(V)=\kappa+Vt_0$. If the voltage is
applied to the edges symmetrically, then $\kappa(V) = \kappa+ V (t_0
+ t_1)/2$. Moreover, if electron tunneling amplitudes are not small,
the current redistribution between the edges due to tunneling
affects the average electron numbers $N_{0,1}$, and the interference
phase $\kappa$ in general should be determined self-consistently. On
the other hand, if the charge density is fixed by electrostatics and
voltage $V$ cannot change the chemical potentials of the two edges,
the phase difference should be independent of $V$: $\kappa(V)=
\kappa$.

The operator of the electron tunnel current from the edge $0$ into
the edge $1$ is found to have the usual form
\[ I^{e}=i[\int dx \rho_0(x),{\cal H}]=\frac{\delta}{\delta
\phi_0}{\cal L}_t=i\sum_{j=1,2}\sum_{\pm} (\pm) T_j^{\pm} e^{\mp
iVt} \, .\]
Its average contains the phase-insensitive contribution $\bar{I}^e$
from the two point contacts independently, and the phase-sensitive
interference term $\Delta I^e(\kappa)$:
\[ I=\langle I^e \rangle =\bar{I}^e+\Delta I^e(\kappa). \]

\subsection{Perturbative calculation of electron tunneling
current}

In the lowest non-vanishing order of the perturbation theory in
$U_j$, the average tunneling current can be calculated as
\begin{eqnarray}
&&I(V)=i\int^0_{-\infty}dt \langle [I^e(0),{\cal L}_t(t)] \rangle
\nonumber \\ &&=\int^\infty_{-\infty}dt e^{iVt} \langle [\sum_j
T^-_j(t),\sum_k T^+_k(0)] \rangle \, , \label{2c1}
\end{eqnarray}
where the average $\langle ... \rangle$ is taken over the states of
the two free propagating edges. Substituting the bosonic expression
from Eq.~(\ref{e2}), one finds the phase-insensitive term consisting
of the two contributions from individual point contacts:
\begin{eqnarray}
\bar{I}^e &=& 2i\sum_j (\frac{DU_j}{2\pi})^2 \int^\infty_{-\infty}
dt e^{\lambda^2(<\varphi_j(t)\varphi_j(0)>-<\varphi_j^2>)} \sin Vt
\nonumber \\ &=& \sum_j(U^2_jD/2\pi) (2\pi T/D)^{\lambda^2-1}
C_{\lambda^2}(V/2\pi T)\, , \label{e3}
\end{eqnarray}
where the second line follows \cite{kf} from Eq.~(\ref{e0}) for the
bosonic correlator, and
\[ C_g (v) \equiv \sinh (\pi v)|\Gamma (g/2 +iv)|^2/ [\pi\Gamma(g)]
\, . \]
For $g$ equal to an even positive number, this function reduces to
the polynomial, $C_g (v)=v \prod_{n=1}^{g/2-1}(n^2+v^2)/\Gamma(g)$.

The interference term can be written as
\begin{eqnarray}
\Delta I^e=(\frac{U_1U_2D^2}{\pi^2}) \int^\infty_{-\infty}dt \
\mbox{Im}\ e^{\lambda^2(<\varphi_2(t)\varphi_1(0)>-<\varphi_1^2>)}
\cdot  \nonumber \\ \sin [ \kappa(V)-Vt ]= (\frac{U_1U_2D}{\pi^2})
(\frac{\pi T}{D})^{\lambda^2 -1}
\int^\infty_{-\infty} ds \sin [ \kappa(V)- \nonumber \\
V\bar{t} -\frac{sV}{\pi T} ] \, \mbox{Im} \Big\{ \prod_{l=0,1}
[i\sinh(s-(-1)^l \Delta t\pi T-i0)]^{-1/\nu_l} \Big\} , \label{e4}
\end{eqnarray}
in the notation $t_{0,1}=\bar{t} \pm \Delta  t$. After
redefinition of the phase,
$\kappa_V=\kappa(V)-V\bar{t}$, Eq.~(\ref{e4}) coincides
with the interference term obtained in the antidot
geometry \cite{gl,b9}. Since the powers $1/\nu_l$ are
integer, the integral in Eq.~(\ref{e4}) can be
transformed into a closed contour integral and evaluated
by residues as follows
\begin{eqnarray}
\Delta I^e= (\frac{U_1U_2D^2}{\pi^2})(\frac{\pi T}{i D})^{\lambda^2}
\sum_{m=0,1} \frac{\pi}{\Gamma(1/\nu_m)}
\partial_s^{1/\nu_m-1} \Big\{ \nonumber \\ \cdot  \frac{(s-(-1)^m \Delta
t)^{1/\nu_m} \sin (sV-\kappa_V)}{ \prod_{l=0,1} [\sinh((s-(-1)^l
\Delta t)\pi T-i0)]^{1/\nu_l}} \Big\}|_{s=(-1)^m \Delta t} \, .
\label{e5}
\end{eqnarray}
At $T=0$, this expression describes an oscillating behavior of the
phase-sensitive current. In the case $V\Delta t \gg 1$, it is
characterized by the asymptotics
\begin{equation*}
\Delta I^e \simeq {2U_1U_2 D \sin(V\Delta t)\cos(\kappa_V)
(V/D)^{1/\nu-1} \over \pi i^{1/\nu+1}(1/\nu-1)!(2D\Delta t)^{1/\nu}}
\end{equation*}
for $\nu_0=\nu_1 \equiv \nu$, and
\begin{equation}
\Delta I^e \simeq {U_1U_2 D \sin(Vt_0-\kappa(V)) (V/D)^{1/\nu_0-1}
\over \pi i^{1/\nu_1+1} (1/\nu_0-1)!(2D\Delta t)^{1/\nu_1}} \ \
\label{e6} \end{equation} for $\nu_0 \neq \nu_1$. The integral
visibility of the interferometer is defined as
\[ Vis \equiv ( \max_\kappa I-\min_\kappa I)/(\max_\kappa I
+\min_\kappa I), \]
where the minimum and maximum are taken over the dependence of the
current on the interference phase $\kappa$. Substituting the
large-$V$ asymptotics of current into this definition, one finds that
for equal filling factors $\nu_0=\nu_1 \equiv \nu$ the visibility
decreases and oscillates with voltage as
\begin{equation}
Vis \simeq{(2/\nu-1)! \over (1/\nu -1)! } {4 U_1U_2 \over
U^2_1+U^2_2} {|\sin(V\Delta t)|\over |2\Delta t V|^{1/\nu}}\, .
\label{e7} \end{equation}
For $\nu=1$, this asymptotics becomes an exact
expression for the integral visibility of integer
quantum Hall MZI {Ref.~\cite{integer}) and was tested in
experiments \cite{mz01}. In this case, suppression of
the interference is caused by linear variation in the
interference phase with energy of propagating electrons.
For $\nu_0<\nu_1$, the oscillations vanish
asymptotically as:
\begin{equation}
Vis \simeq{(\lambda^2-1)! \over (1/\nu_0-1)! } {4 U_1U_2
\over U^2_1+U^2_2} |2\Delta t V|^{-1/\nu_1}\, .
\label{e8}
\end{equation}
Both expressions (\ref{e7}) and (\ref{e8}) generalize the
description of the suppression of the interference due to variation
of the phase of propagating excitations with energy from the case of
integer edges to the edges with fractional filling factors.

In the opposite limit of $V,T< 1/\Delta t$, the right-hand-side of
Eq.~(\ref{e5}) sums up to the same polynomial $C_{\lambda^2}(V/2\pi
T)$, and the full current $\langle I^e \rangle$ is given by
Eq.~(\ref{e3}) with the sum of squares of the two point-contact
amplitudes $U_j$ replaced by square of their coherent sum:
\begin{equation}
I= \frac{|U_1+U_2e^{i\kappa_V }|^2D}{2\pi}(2\pi
T/D)^{\lambda^2-1} C_{\lambda^2}(V/2\pi T)\, .
\label{e9}
\end{equation}
In this regime, the visibility reaches its maximum
\[ Vis=2U_1U_2/(U_1^2+U_2^2). \]
Appearance of the geometric sum of the two tunneling
amplitudes in Eq.~(\ref{e9}) suggests that for small
$\Delta t$, the two-point-contact model of MZI described
with Lagrangian (\ref{e2}) reduces to a
single-point-contact model, but with the new amplitude.
Indeed, such a single-point-contact model would provide
an appropriate equivalent description of the two-point
tunneling of non-interacting electrons, e.g., in a
single-mode edge states in integer quantum Hall effect.
Therefore, this reduction agrees with the usual practice
\cite{kf} in studies of one-dimensional interacting
electrons in quantum wires, where the two operations:
the low-energy reduction of the multiple electron
scattering to an effective single scatterer, and the
switching on of the electron-electron interaction, are
treated as interchangeable. However, the problem of FQHE
edge states transport is different. In particular, the
dynamics of FQHE edge states is affected not by 1D but
2D electron interaction and geometry. We will show below
that the interchange of the order of the operations is
valid only in the lowest order in the limit of weak
tunneling. In general, it would contradict the following
physical feature of the system. The
weak-electron-tunneling description (\ref{e2}) of the
interferometer is intrinsically related to a
strong-tunneling model. Indeed, as follows from
Eq.~(\ref{e9}), both of the amplitudes $U_j$ scale at
low energies $E$ roughly as $E^{\lambda^2/2-1}$, where
$E\simeq \mbox{max} (V,T)$, and therefore increase with
energy. The model of two FQHLs strongly coupled at two
point contacts separated by a finite $\bar{t}$
possesses, however, a different topology than the
single-point-contact model. Since the FQHL is a
topological quantum liquid \cite{b12}, topology with two
coupled points implies multiple degeneracy of the ground
state, which leads, as will be seen more explicitly
below, to the tunneling current different from that in
the single point contact.

\section{Strong-coupling limit}

\subsection{Quantum nature of the effective magnetic flux}

To derive the dual strong-coupling model for the MZI at large
effective tunneling amplitudes $U_j$, we treat the problem in
imaginary time and use the standard instanton technique. The ground
states are determined by minimization of the action ${\cal S}$:
\begin{equation}
{\cal S}={\cal S}_{kin}+{\cal S}_{t} \, , \label{action}
\end{equation}
which includes the tunneling part ${\cal S}_{t}$ defined
by Lagrangian (\ref{e2}) and the kinetic term ${\cal
S}_{kin}$ defined by Eqs.~(\ref{e1}) and
(\ref{aftere2}). In the limit $U_j\gg 1$ for both
$j=1,2$, the tunneling Lagrangian (\ref{e2}) gives the
dominant contribution to the action in Eq.
(\ref{action}). If the two parts of the Lagrangian
(\ref{e2}) that describe the two contacts were treated
separately, both tunneling modes $\varphi_j$, $j=1,2$,
would be fixed at the extrema of the corresponding parts
of the Lagrangian. Considering both $T_j^{\pm}$
together, one can see, however, that their equal-time
interchange relation is
\begin{equation}
T_2^{\pm}T_1^{\mp}=e^{2\pi m i} T_1^{\mp} T_2^{\pm} \, , \label{e10}
\end{equation}
as follows from the commutation relation
$[\varphi_2,\varphi_1]=i \pi$. As discussed in more
details below, this relation represents the essence of
the MZI interference physics. It shows that although the
different transfer terms $T_j^{\pm}$ [Eq.~(\ref{e2})]
commute among themselves, each interchange of the
electron tunneling processes at the two contacts changes
the interference phase $\kappa$ so that the external
magnetic flux $\Phi_{ex}$ acquires an additional
contribution $\pm m\Phi_0$. This mechanism transforms
the external flux into effective flux $\Phi$ which
includes the statistical contribution, as discussed in
Section I. While this statistical flux is irrelevant in
the situation of weak electron tunneling, it becomes
crucial for the quasiparticle tunneling, when
effectively one needs to split the electron transfer
terms with the associated exchange phase (\ref{e10})
into the transfer terms for the quasiparticles with
fractional charge $e/m$. The corresponding splitting of
the exchange phase into $2\pi/m$ terms is non-trivial.

\subsection{Ground states and bosonization of Klein factors}

The new statistical flux mechanism discussed above should affect the
construction of the ground states of the interferometer in the
strong-tunneling regime. To see how this happens, we first examine
the perturbative expansion of the partition function in $U_{1,2}$.
When imaginary times of two tunneling processes at different points,
$T_{1}^{\pm}$ and $T_{2}^{\mp}$, change their time order, the phase
branch of the perturbative term changes accordingly to
Eq.~(\ref{e10}). In general, one can make different choices for the
phase branches by multiplying the tunneling operators $T_{j}^{\pm}$
with some Klein factors $\exp\{\pm i\sqrt{2\gamma} \eta_j\}$, where
the free zero-energy bosonic modes $\eta_j$ are defined by their
imaginary-time-ordered correlators:
\begin{equation} \langle T_\tau{\eta_i(\tau) \eta_j(0)}\rangle =i \pi
\Theta((j-i)\tau) (1-\delta_{ij}).
\label{stat} \end{equation}
For any integer $\gamma$, incorporation of these Klein
factors into the terms $T_j^{\pm}$ in Eq.~(\ref{e2})
does not change the perturbation expansion of the
partition function in ${\cal S}_t$ in any order. Even
integer $\gamma$ affects, however, the kinetic part of
the action. As we show below, this fact can be used to
construct the ground states which minimize the energy of
the system in the strong-coupling limit.

Indeed, the new tunneling fields $\Phi_j$ which include
the modes $\eta_j$:
\[ \Phi_j=\lambda \varphi_j+ \sqrt{2\gamma} \eta_j\]
are characterized by the kinetic action
\begin{eqnarray}
{\cal S}_{kin}(\gamma,\{\Phi_{1,2}\})=\!\!\int {d \omega \over 4\pi}
\sum_{i,j}(\Phi_i(-\omega) \hat{K}_{ij}^{-1}(\gamma, \omega)
\Phi_j(\omega)\!\!&)&
\label{e11} \\
\hat{K}(\gamma, \omega)=\lambda^2 g(0,\omega) \hat{1}+
\sum_\pm [\mp {2 \pi \gamma \over \omega} +\sum_j
{1\over \nu_j}g(\mp t_j,\omega)\!\!&]&\!\!
\hat{\sigma}_\pm \ , \ \nonumber
\end{eqnarray}
where $\hat{\sigma}_\pm$ are the raising and lowering $2 \times 2$
matrices, and the matrix $\hat{K}(\gamma)$ contains the correlators
\begin{equation}
K_{ij}(\gamma, \tau)=\lambda^2 \langle T_\tau\varphi_i(\tau)
\varphi_j(0)\rangle + 2\gamma \langle T_\tau\eta_i(\tau) \eta_j(0)
\rangle \, . \nonumber
\end{equation}

Next, to construct the ground states, we follow the
procedure from Ref.~\cite{us} and express the energy
$E_{t}$ associated with the electron tunneling
Lagrangian (\ref{e2}) in terms of the low-temperature
asymptotics $(\beta \equiv 1/T \to \infty)$ of the
partition function of the system:
\begin{equation}
e^{-\beta E_{t}}=\frac{\int D\Phi e^{-{\cal
S}(\gamma,\{\Phi_{1,2}\})}} {\int D\Phi e^{-{\cal
S}_{kin}(\gamma,\{\Phi_{1,2}\})}} \, . \label{partition}
\end{equation}
Here the integrations $D\Phi \equiv \prod_{j=1,2}D
\Phi_j(\tau)$ run over functions defined on the
imaginary time interval $ \tau \in [0, \beta]$ with the
periodic boundary conditions. According to
Eq.~(\ref{action}), the action ${\cal
S}(\gamma,\{\Phi_{1,2}\})$ consists of the kinetic term
${\cal S}_{kin}(\gamma,\{\Phi_{1,2}\})$
[Eq.~(\ref{e11})] and the tunneling part ${\cal S}_t$,
which after substitution of the Klein factors takes the
following form:
\begin{equation*}
{\cal S}_t(\{\Phi_{1,2}\})=-\int_0^\beta d\tau \sum_{j=1,2} {D U_j
\over \pi} \cos(\Phi_j+\kappa_j) \, .
\end{equation*}
As discussed above, in the limit of small $U_{1,2}$,
where the perturbative expansion in the electron
transfer terms $T_j$ is applicable, the energy $E_{t}$
in EQ.~(\ref{partition}) does not depend on $\gamma$.
However, in the strong-coupling limit of large $U$'s,
the dominant tunneling part of the action ${\cal S}$
imposes the strong-tunneling conditions:
\begin{equation}
\Phi_j=2 \pi n_j-\kappa_j\equiv \Phi_{n_j} , \;\;\; \mbox{for} \
j=1,2, \ \label{e12}
\end{equation}
in the upper functional integral in Eq.~(\ref{partition}). As a
result, the energy $E_{t}$, which can be expressed as
\begin{equation}
E_{t}=\int\! {d\omega \over 4 \pi}  \ln[\mathsf{Det}
\hat{K}(\gamma,\omega)] e^{-|\omega|/D}+\frac{{\cal
S}(\gamma,\{\Phi_{n_j}\})}{\beta} \, , \label{etunn} \end{equation}
acquires  dependence on the parameter $\gamma$.
Substitution of the matrix $\hat{K}(\gamma)$ from
Eq.~(\ref{e11}) into Eq.~(\ref{etunn}) gives the
$\gamma$-dependent part of $E_{t}$ as
\begin{eqnarray*}
E_{t}\!\!&=&\!\!\int\!\! {d\omega \over 4 \pi} \
\ln[m^2+(\gamma-m)^2+
(\gamma-m)\sum_j \frac{e^{-|\omega|t_j}}{\nu_j}] \\
&\times& e^{-|\omega|/D} +{\delta_{\gamma,0}
(n_1-n_2+\kappa)^2\over \sum_j (t_j/\nu_j)} +const \, .
\end{eqnarray*}
Minimization of this expression at $t_{0,1}D \gg 1$
imposes unambiguously the choice $\gamma=m$, which also
guarantees commutativity of the two tunneling fields
$\Phi_j$. The commutativity is important to make the
strong-tunneling conditions [Eq.~(\ref{e12})]
self-consistent. On the other hand, in the limit
$t_{0,1}D \to 0$, the minimum of $E_{t}$ occurs for
$\gamma=0$. This is precisely the limit when we can be
sure that the tunneling at the two point contacts
reduces to a single-point tunneling characterized by the
effective amplitude equal to the geometrical sum of the
two point-contact amplitudes.

Next, we discuss briefly how the incorporation of the
chosen bosonic Klein factors into the full tunneling
operators $T_{Cj}^\pm \equiv T_{j}^\pm \exp\{\pm i
\sqrt{2m} \eta_j\}$ affects our earlier interpretation
of the physics underlying Eq.~(\ref{e10}).
Qualitatively, the Klein factors change the dynamics of
the interchange relations Eq.~(\ref{e10}). Indeed, one
can see directly that the phase factor of
Eq.~(\ref{e10}) drops out from the equal-time
interchange relation of the full tunneling operators
$T_{Cj}$. It, however, re-appears when their time
difference is much larger than both propagation times
$t_{0,1}$. In particular, we find that for $\tau \gg
t_{0,1}$
\begin{equation}
(T_{C2}^-T_{C1}^+)(0)T_{C1,2}^+(-\tau)=e^{-2\pi m i}
T_{C1,2}^+(\tau)(T_{C2}^-T_{C1}^+)(0) \, . \label{e10new}
\end{equation}
This new relation characterizes dynamics of the mechanism of the
effective flux transformation we discussed above: Each process of
electron tunneling in any of the two contacts of the interferometer
modifies the interference pattern for subsequent electrons passing
through the interferometer at much later times, as if the effective
flux is changed by $m \Phi_0$.

\subsection{Instanton expansion and duality transformation}

The standard instanton calculation of the partition
function ${\cal Z}$ for infinitely degenerate series
(\ref{e12}) of the ground states $(\Phi_{n_1},
\Phi_{n_2})$ leads us to the expression ${\cal
Z}=\sum_{n_j} {\cal Z}_{n_1,n_2}$. Each term in this sum
is calculated through the substitution into
$\exp\{-{\cal S}(\Phi_1,\Phi_2)\}$ of the asymptotic
form of the instanton expansion around the $(\Phi_{n_1},
\Phi_{n_2})$ ground state:
\[ \Phi_j(\tau)=\Phi_{n_j}+\sum_l 2\pi e_{l,j} \theta
(\tau-\tau_{l,j})\, , \] and further summation over the number of
instantons $e_{l,j}=1$ and anti-instantons $e_{l,j}=-1$, and
integration over their times $\tau_{l,j}$. The result can be
presented in the following form:
\begin{eqnarray}
&   & {\cal Z}_{n_1,n_2}\propto \int  D\Theta_{1,2} \exp\{-{\cal
S}^D_{kin}+\sum_j\!{W_j D\over 2 \pi} \cdot \nonumber \\
& & \int d\tau \cos \large[ \Theta_j(\tau) +(-1)^j (\kappa_j-2 \pi
n_j)/ m \large] \} \label{e14}
\end{eqnarray}
with a constant of proportionality independent of $n_{1,2}$. The new
kinetic term in this action is defined as
\begin{eqnarray}
{\cal S}^D_{kin}(\Theta)={1 \over 2} \int {d \omega \over 2\pi}
\Theta(-\omega) [(2\pi/\omega)^2\hat{K}^{-1}(\omega)]^{-1}
\Theta(\omega)\, , \;\; \label{e15} \\
\big({2\pi \over \omega}\big)^2\hat{K}^{-1}={4 \pi \over \lambda^2
|\omega|}\hat{1}+ \sum_{\pm,j} \pm {8 \pi\over \lambda^4
\omega}{e^{\pm \omega t_j}\over \nu_j}\theta(\mp\omega)
\hat{\sigma}_\pm \, , \;\;\;  \nonumber
\end{eqnarray}
by instanton-instanton interaction, while the phases of cosines in
Eq.~(\ref{e14}) follow from the interaction between instantons and
the $n_{1,2}$ ground state.

Comparing the correlators of the fields $\Theta$ defined by this
action to $g(z,\omega)$ in Eq.~(\ref{e1}), we can divide these
fields into the two parts:
\begin{equation}
\Theta_j=(-1)^j [\sqrt{2\over m} \eta_j+ {2 \over
\lambda}\vartheta_{j}]\ . \label{e16}
\end{equation}
The bosonic modes $\eta_{1,2}$ describe here purely
statistical effect (\ref{stat}), while the fields
$\vartheta$ have the chiral correlators:
\begin{equation}
\langle \vartheta_{j}^2 \rangle =g(0,\omega)\, ,\ \langle
\vartheta_{2} \vartheta_{1}\rangle = {g(t_0,\omega)\over
\nu_0\lambda^2}+{g(t_1,\omega)\over \nu_1\lambda^2}\, . \label{e17}
\end{equation}
Notice further that the contribution ${\cal
Z}_{n_1,n_2}$ in Eq.~(\ref{e14}) depends on both $n_1$
and $n_2$ only through their difference modulo $m$.
Therefore, up to a divergent constant, ${\cal Z}$
becomes a finite sum. This sum over the indices combined
with integration of the exponents of the re-extracted
statistical fields can be represented as a trace over
the $m$-dimensional Hilbert space. This is achieved by
ascribing to each instanton tunneling exponent in
Eq.~(\ref{e14}) a proper $m$-dimensional matrix. These
unitary matrices $\bar{F}_j$ are characterized by the
following relations:
\begin{equation}
\bar{F}_1\bar{F}_2=e^{-2 \pi i\over m}
\bar{F}_2\bar{F}_1 , \;\;\; \langle \bar{F}^k_1
(\bar{F}^+_1)^p\bar{F}^l_2 (\bar{F}^+_2)^q \rangle
=\delta_{kp}\delta_{lq}\, , \label{e17next}
\end{equation}
where the Kronecker symbol $\delta_{ij}$ is defined
modulo $m$. The first relation in Eq.~(\ref{e17next}) is
due to the statistical parts of the fields $\Theta$ in
Eq.~(\ref{e16}), while the second one follows from the
$m$-periodic dependence of Eq.~(\ref{e14}) on both
indices. Writing ${\cal Z}$ in the form of the trace
makes it equal to a partition function of the
quasiparticles whose tunneling Lagrangian $\bar{{\cal
L}}_t$ in real time has the form dual to the Lagrangian
(\ref{e2}):
\begin{eqnarray}
\bar{{\cal L}}_t = \sum_{j=1,2} \Big[ {W_jD\over 2\pi } \bar{F}_j
\exp \big\{ i \big({\kappa_j(V) \over m} +{2 \vartheta_j \over
\lambda} - {Vt \over m}\big) \big\} \nonumber \\ + h.c.\Big] \equiv
\sum_{j=1,2} \sum_{\pm} \bar{T}_j^{\pm} e^{\mp iVt /m} . \label{e18}
\end{eqnarray}
The operators $\bar{F}_j$ here are the Klein factors
describing the statistics of the quasiparticles. These
factors are analogous to the Klein factors derived in
Ref.~\cite{b9} for the quasiparticle tunneling in the
antidot geometry. They act in the Hilbert space spanned
by the $m$-fold degenerate ground state of the MZI in
the absence of the quasiparticle tunneling. As discussed
in Section I, in both the antidot and the MZI
geometries, the $m$ states correspond to different
effective flux $\Phi$ enclosed by the edges between the
two point contacts. The quasiparticle model of the MZI
based on the tunnel Lagrangian (\ref{e18}) derived above
generalizes the quasiparticle model in Ref.~\cite{mz3}
which used a particular form of the matrix Klein factors
complying with Eq. (\ref{e17next}) up to a phase factor
we find below.

Finally, using the quasiparticle expression
(\ref{qpcurrent}) for the tunneling current that is
obtained in Section III D, we proceed to the
perturbative calculation of this current
\begin{eqnarray}
&&I(V)=i\int^0_{-\infty}dt\langle [I^q(0),\bar{\cal L}_t(t)] \rangle
\nonumber \\ &&={1\over m}\int^\infty_{-\infty}dt e^{iVt/m} \langle
[\sum_j \bar{T}^-_j(t),\sum_l \bar{T}^+_l(0)] \rangle  \, .
\label{3c1} \end{eqnarray}
It should be noted that the average in this expression
includes, in particular, the average over the
$m$-dimensional Hilbert space of the flux states taken
according to Eq.~(\ref{e17next}). This makes the
interference term vanish in the lowest perturbative
order, the fact that suggests suppression of the
interference in general in the model (\ref{e18}) of the
quasiparticle tunneling. In our discussion of the
electron tunneling model in Sec.~II, we saw, however,
that the interference is suppressed only if $V\Delta t
\gg 1$ or $T\Delta t \gg 1$. As will be shown below, the
same is true in the regime of the quasiparticle
tunneling. Validity of the perturbative result
(\ref{3c1}) which predicts suppressed interference is
indeed restricted to the regimes of $V\Delta t \gg 1$ or
$T\Delta t \gg 1$. For $(V,T) \Delta t < 1$, solution of
Lagrangian (\ref{e18}) is non-perturbative.

\subsection{Boundary conditions, dual chiral fields, and edge
currents}

To clarify the dynamics of the tunneling fields
$\vartheta_j$ defined by Eq.~(\ref{e16}), and to explain
the introduction of the applied voltage in
Eq.~(\ref{e18}), we need to relate the tunneling fields
$\vartheta_j$ to the incoming edge modes $\phi_{0,1}$.
To do this, we first consider the case of $\Delta t=0$
and equal velocities of the edge modes $\phi_l$.  In
this case, both tunneling bosonic operators $\varphi_j$
in Eq.~(\ref{e2}) are just the two operator values of
the same bosonic field $\phi_-$ at points $x_{1,2}$:
$\varphi_j= \phi_-(x_j)$, where in the absence of
tunneling, the field
\begin{equation}
\phi_-={\sqrt{\nu_1}\phi_0-\sqrt{\nu_0}\phi_1 \over \sqrt{\nu_0 +
\nu_1}} \label{phi-} \end{equation}
is a free chiral filed. The combination $\phi_+$ of the two edge
modes that is orthogonal to $\phi_-$:
\begin{equation}
\phi_+={\sqrt{\nu_0}\phi_0+\sqrt{\nu_1}\phi_1 \over \sqrt{ \nu_0 +
\nu_1}}\, , \label{phi+} \end{equation}
is not affected by tunneling at all and is always a free chiral
field. In the strong-coupling limit of the two tunneling terms
\[ {\cal L}_{t,j}=(D U_j/\pi) \cos \large[ \lambda \phi_-(x_j)+
\kappa_j \large] \]
treated independently of each other, the propagation of $\phi_-$ is
described by imposing the Dirichlet boundary condition. "Unfolded"
form of this condition \cite{bc} implies free chiral propagation of
the fields $\mbox{sgn}(x-x_j)(\phi_-(x)+\kappa_j/\lambda)$ across
each point contact $x_j$. Application of these boundary conditions
at both contacts successively implies that the field outgoing from
the first contact is used as the incoming filed for the second
contact. This procedure results in the free propagation of the
chiral field that can be written for all values of $x$ as
\begin{eqnarray}
\vartheta_-(x)\!\!&=&\!\!\phi_-(x)\theta(x_1-x)+(\phi_-(x)-2 {\kappa
\over \lambda})\theta(x-x_2) \nonumber \\ &-&\!\!(\phi_-(x)+2
{\kappa_1 \over \lambda})\theta(x-x_1) \theta(x_2-x)\, . \label{e19}
\end{eqnarray}
This strong-coupling propagation of $\phi_-(x)$ implies
that it changes sign and acquires some phase shifts on
both passages through $x_j$. Finite quasiparticle
backscattering leads to deviations from the free chiral
propagation (\ref{e19}) and is described by the dual
tunneling terms $\bar{\cal L}_{t,j}=(D W_j/\pi)
\cos[(2/\lambda )(\phi_-(x_j)+\kappa_j/\lambda )]$.
Expressed through the free chiral dual field
$\vartheta_-$ (\ref{e19}), these dual tunneling terms
take the form
\[ \bar{\cal L}_{t,j}=(D W_j/\pi) \cos \large[ (2/\lambda)
(\vartheta_-(x_j)+\kappa_j/\lambda ) \large]. \]
Their comparison with the tunneling Lagrangian in
Eq.~(\ref{e18}) derived by the instanton expansion shows
that the tunneling fields $\vartheta_j$ in
Eq.~(\ref{e18}) are related to the dual chiral field as
$\vartheta_j=\vartheta_-(x_j)$ in agreement with
Eq.~(\ref{e17}). Both parts $\sum_{\pm} \bar{T}_{1,2}
^{\pm}$ of the tunneling Lagrangian (\ref{e18}) are
constructed from the dual tunneling fields which are
combined with the Klein factors to restore the
commutativity.

To understand how the applied voltage $V$ enters in
(\ref{e18}), we note that the applied voltage can be
introduced at first as a shift of the incoming field:
$\phi_0-\sqrt{\nu_0}Vt$. As one can see from
Eqs.~(\ref{phi-}) and (\ref{e19}), this shift translates
into the shift $\vartheta_--Vt/\lambda$ of the dual
field, producing the voltage bias in the quasiparticle
Lagrangian shown in Eq.~(\ref{e18}). Also, since at the
end of its propagation through the MZI, the $\phi_-$
field again coincides with the $\vartheta_-$ field
(\ref{e18}) up to a constant, the tunneling current in
the MZI is produced only by the deviations of the
$\vartheta_-$ field from its free propagation that are
caused by the dual tunneling terms. Relating variations
in $\phi_0$ to variations in $\phi_-$ through
Eqs.~(\ref{phi-}) and (\ref{phi+}), one finds the
quasiparticle tunneling current to be equal to
\begin{equation}
I^q={i\over \lambda} [\bar{{\cal L}}_t, \int {dx \over 2 \pi}
\partial_x \vartheta_-(x)]={i\over m} \sum_{j=1,2}\sum_{\pm} \pm
\bar{T}_j^{\pm} e^{\mp iVt/m} \, . \label{qpcurrent}
\end{equation}
Both of these results, for the bias voltage and for the tunneling
current, can be understood simply as manifestations of the
fractional charge $e/m$ of the quasiparticles.

\begin{figure}[htb]
\setlength{\unitlength}{1.0in}
\begin{picture}(3.3,.55)
\put(0.2,-0.1){\epsfxsize=2.8in \epsfbox{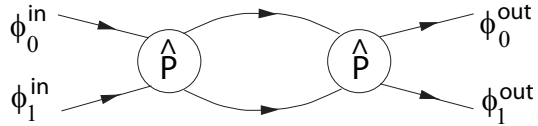}}
\end{picture}
\caption{Diagram of the strong-coupling edge propagation in the
Mach-Zehnder interferometer.}
\end{figure}

The picture of successive splitting of the edges at the two point
contacts in the strong-coupling regime that underlies
Eq.~(\ref{e19}) remains valid for $\Delta t \ne 0$, with the edge
fractions propagating along the two sides of the interferometer
remaining the same as those that follow from Eqs.~(\ref{phi-}),
(\ref{phi+}), and (\ref{e19}). This means that the whole picture of
propagation of charges and currents carried by the $\phi_j$ fields
in the MZI can be represented in general with the diagram shown in
Fig.~2, where each of the matrices $\hat{P}$ is \cite{b14,b8}
\begin{equation}
P_{00}=-P_{11} = {\nu_0- \nu_1\over \nu_0+ \nu_1} \, ; \;\;
P_{01}=P_{10} = -{2\sqrt{\nu_0 \nu_1} \over \nu_0 +\nu_1} \, ,
\label{e20} \end{equation}
and describes the edge splitting at the point contact.
The matrix (\ref{e20}) satisfies the identity $\hat{\bf
P}^2=\hat{\bf 1}$, which implies that the two
consecutive scattering processes at the point contacts
do not change the current distribution between the modes
of the interferometer. For instance, in the case of
equal filling factors, $\nu_0=\nu_1$, the matrix
$\hat{\bf P}$ just interchanges the edge modes in both
contacts. Therefore, the tunneling current in the MZI is
created only by the dual tunneling terms, and
Eq.~(\ref{qpcurrent}) for this current is valid
independently of $\Delta t$.

Besides transferring fractional charge of quasiparticles $e/m$, each
dual tunneling also changes the effective flux through the MZI. This
can be seen explicitly from the relation
\begin{equation}
(\bar{T}_{2}^-\bar{T}_{1}^+)(0)\bar{F}_{1,2}(-\tau)=e^{-2 \pi i\over
m} \bar{F}_{1,2}(\tau)(\bar{T}_{2}^-\bar{T}_{1}^+)(0) \, ,
\label{e20next} \end{equation}
which  follows from Eq.~(\ref{e17next}). For $\tau \gg
t_{0,1}$, the bosonic exponents of the tunneling
operators do not affect this interchange relation.
Equation (\ref{e20next}) is analogous to
Eq.~(\ref{e10new}) for electrons, and characterizes
dynamics of the effective flux transformation: Each
tunneling of a quasiparticle in any of the two
interferometer contacts adds flux quantum $\Phi_0$ to
the effective flux through the interferometer, which
modifies the interference phase for quasiparticles
tunneling much later through the interferometer. In this
respect, the interchange relations (\ref{e20next}) agree
with the physical picture suggested in Refs.
~\cite{mz4,mz3} and derived earlier in the context of
the antidot tunneling \cite{b9}, in which each tunneling
quasiparticle also carries with it a flux quantum. An
important unresolved question of this picture is to what
extent the statistical contribution to the interference
phase can be understood directly as a real change in the
magnetic flux through the interferometer.

The simplest $m$-dimensional irreducible representation for the
Klein factors that account for this flux changes is:
$$X_{l,j}=\delta_{l+1,j} \ (\mbox{mod}\ m), \ Y_{l,j}=\delta_{l,j}
e^{-i{2\pi \over m}l} \ .$$  If we apply a unitary transformation,
it takes a flux-diagonal form:
$$\bar{F}_1=X, \bar{F}_2=-e^{\pm i\pi/m}XY \ .$$ The phase
factor here follows from the second equation in
Eq.~(\ref{e17next}). In the case $m=2$, these Klein
factors are equal to the Pauli matrices:
$\bar{F}_{1,2}=\sigma_{X,Y} $.

\section{Shot noise of the tunneling current}

In general, the noise power spectrum of the current $I^a$ at
frequency $\omega$ and voltage $V$, where $a=e,q$ denotes,
respectively, the electron or quasiparticle form of the tunneling
current, is defined as
\[ P(\omega,V)=\int_{-\infty}^\infty dt \cos(\omega t) (\langle
I^a(t)I^a(0)\rangle -\langle I^a \rangle^2) \, . \]
In the lowest-order perturbation theory in the respective tunneling
amplitudes, this expression takes form
\begin{eqnarray}
P(\omega,V)&=&q_a^2\int^\infty_{-\infty}dt\cos( \omega t)e^{iVt q_a}
\nonumber \\ &\times& \langle \{\sum_j T^{a -}_j(t),\sum_l T^{a
+}_l(0)\} \rangle \, . \label{n1}
\end{eqnarray}
Here $q_a$ is the charge of the tunneling particles which is equal
to $1$ or $1/m$ in units of the electron charge $e$, and $T^{a\pm}$
stands for $T^{\pm}$ or $\bar{T}^{\pm}$ in the regime of,
respectively, electron and quasiparticle tunneling.  Comparison of
Eq.~(\ref{n1}) with perturbative expressions for the tunneling
current in Eqs.~(\ref{2c1}) and (\ref{3c1}) shows immediately that
the average currents and the noise power spectra are related in the
lowest order of the perturbation theory as
\begin{equation}
P(\omega,V)={q_a \over 2} \sum_\pm \coth[(q_a V \pm \omega)/(2T)]
I(q_aV \pm \omega) \, . \label{n2}
\end{equation}

On the other hand, the zero-frequency limit of the noise
power and the average current can be found from the
long-time asymptotics of the distribution of the
tunneling charge as its second and first moment divided
by time.  At $T=0$, and in the lowest order in the
tunneling amplitudes, this distribution corresponds to a
Poisson process, and therefore the Schottky formula for
the shot noise $P(0,V)/I(V)=q_a$ displays directly the
charge of the tunneling particles. This charge is $e$ at
low voltages, and $e/m$ at sufficiently large voltages,
if $V\Delta t>1$ and the perturbative treatment of the
quasiparticle tunneling model is correct. It is
important to stress here that although the fractional
charge $e/m$ coincides with the charge $e_X$ of the
point-contact quasiparticles in the situation of
tunneling in one point contact \cite{b14}, it appears in
Eq.~(\ref{n2}) for the quasiparticle noise in the MZI by
a purely statistical mechanism. This mechanism is the
reduction in the MZI flux variations from $\pm m
\Phi_0$, which are produced by electron tunneling in
accordance with the composite-fermion statistics of
electrons in both edges, to $\pm \Phi_0$ associated with
the quasiparticle tunneling. This interpretation of the
statistical origin of the fractional charge in
Eq.~(\ref{n2}) is supported also by the analysis
\cite{b9} of the antidot interferometer, where the two
charges are different, $e/m \neq e_X$, and $e/m$ is the
charge in the noise spectrum.

\section{Exact solution for symmetric MZI model}

\subsection{Fermionization for $\nu_0=1/3$ and $\nu_1=1$}

We now return to the case of symmetric interferometer with $\Delta
t=0$, and consider the derived dual model (\ref{e18}) in the case
$\lambda=2$ (i.e., for $\nu_0=1/3$ and $\nu_1=1$), when the main
parts, $e^{\pm i \vartheta_-}$, of the quasiparticle tunneling
operators have the same correlators as free electrons, and therefore
allow fermionization. Indeed, the Klein factors for $m=2$ can be
represented by two Pauli matrices and fermionized as $\bar{F}_j=i
\xi_j \xi_0$ in terms of the three Majorana fermions $\{\xi_n,
\xi_{n'} \}_+=2 \delta_{n,n'}$. Introducing a chiral fermion field as
\[ \psi=\xi_0 \sqrt{D/(2 \pi v)}e^{i \vartheta_-} , \]
we come to the Hamiltonian
\begin{equation}
{\cal H}=\{{v\over i}\!\! \int\!\! dx\, \psi^+ \partial_x \psi\} -
\sqrt{Dv \over 2 \pi}[\sum_j W_j i \xi_j \psi(x_j) e^{{i\over
2}\kappa_{j,V}}+h.c.] \, , \label{e21} \end{equation}
where the applied voltage is accounted for by the
fermion chemical potential equal to $V/2$. In the
fermionic Hamiltonian (\ref{e21}), the two terms
accounting for successive tunneling at $x_1$ and $x_2$
contain two different Majorana fermions, the fact that
distinguishes this Hamiltonian from the fermionic
Hamiltonian in Ref.~\cite{mz2}. As a result, the
Heisenberg equations of motion which describe scattering
of the field $\psi(x,t)$ at the two point contacts have
the form of the \emph{disentangled} matching conditions
\begin{eqnarray}
i\psi(x)|^{x_j+0}_{x_j-0} = w_j \xi_j \, , \;\;\; w_j\equiv
i\sqrt{D/(2 \pi v)} W_j e^{-{i\over 2}\kappa_{jV}} , \nonumber \\
\partial_t \xi_j(t) = 2iv [w_j\psi^+(x_j,t)-w_j^*\psi(x_j,t)] \, .
\label{e22}
\end{eqnarray}
Free chiral propagation of the field $\psi(x,t)$
everywhere else (away from the point contacts  $x_1$ and
$x_2$) makes it convenient to formulate the scattering
conditions (\ref{e22}) in terms of chiral momentum
eigenstates of the Fourier components $\psi_k$ of the
free chiral field,
\[ \psi_0(x,t)=\int  {d k \over 2 \pi} \psi_k e^{ik(x-vt)}\ , \]
since this field coincide with the field $\psi(x,t)$ in
the absence of scattering. Conditions (Eq.~\ref{e22})
then mix together the components $\psi_k$ and
$\psi^+_{-k}$, which have the same time dependence and
can be interpreted as annihilation operators of particle
and hole states, respectively. Solution of
Eqs.~(\ref{e22}) for scattering at each of the two point
contacts shows that the evolution of the amplitudes of
$\psi_k,\, \psi^+_{-k}$ across this contact can be
described explicitly by the $(2 \times 2)$ scattering
matrix $\hat{\cal S}_{j,k}$ with the elements
\begin{equation}
{\cal S}^{\pm \pm}_{j,k}={k \over k+i2 |w_j|^2}\, , \; {\cal S}^{-
+}_{j,k}={2iw_j^2\over k+i2 |w_j|^2}={\cal S}^{+ - *}_{j,-k}\, .
\label{e23} \end{equation}

Successive scattering of the particles and holes at the two point
contacts is governed then by the scattering matrix $\hat{\cal S}_k$
equal to the product $\hat{\cal S}_k=\hat{\cal S}_{2,k} \hat{\cal
S}_{1,k}$ of the scattering matrices at the two contacts. The
particles incident on the point contacts have the fermi distribution
$f(k,\mu)$ over the momenta $k$ with the chemical potential $\mu=
q_a V$, where the quasiparticle charge is $q_a=1/2$ in units of
electron charge $e$. Summing the scattering processes for particles
with different momenta and using the standard properties of the
scattering matrix,
\[ 1- |S^{++}_k|^2=|S^{-+}_k|^2, \]
we can express the average tunneling current in terms of this matrix
as follows
\begin{eqnarray}
I={1 \over 2} \int { dk \over2 \pi} [ f(k,V
/2)-f(k,-V/2)] |S^{-+}_k|^2 \nonumber
\\ = \int { dk \over2 \pi} [ f(k,V /2)-f(k,0)] {|2ik \sum_j w_j^2|^2
\over \prod_j |(k+2i|w_j|^2)|^2} \, . \label{e24}
\end{eqnarray}
Splitting the product over $j=1,2$ in Eq.~(\ref{e24}) into a
difference of two fractions, and introducing the tunneling rates
$\Gamma_j \equiv 2v|w_j|^2 = DW_j^2/\pi$, one can see that the
current (\ref{e24}) can be expressed as the difference between the
tunneling currents in two individual point contacts:
\begin{equation}
I={|\Gamma_1e^{i\kappa_V}+\Gamma_2|^2 \over \Gamma_1^2 -\Gamma_2^2}
[ I_{1/2}(V,\Gamma_2)-I_{1/2}(V,\Gamma_1) ] \label{e25}
\end{equation}
The tunneling current in a separate point contact is known to be
equal to \cite{kf}:
\[ I_{1/2}(V,\Gamma)=\frac{\sigma_0}{2} [V -2\Gamma \arctan(V/
2\Gamma)]. \]
at vanishing temperature $T$, and to (\cite{t12}):
\begin{equation}
I_{1/2}(V,\Gamma)=\frac{\sigma_0}{2} [\, V -2\Gamma \, \mbox{Im} \,
\psi (\frac{1}{2}+\frac{2\Gamma+iV}{2\pi T})]. \label{4a10}
\end{equation}
at non-zero $T$. Here $\psi (z) = d\ln \Gamma (z)/dz$ is the digamma
function, and $\sigma_0=e^2/2\pi \hbar$ is the conductance quantum
equal to $1/2\pi$ in the units ($e=\hbar=1$) used in this paper.

At $T=0$, the low-voltage asymptotics of the tunneling current $I$
in the MZI is proportional to $V^3$ and coincides with the electron
tunneling current in Eq.~(\ref{e9}) under the condition $U_j=\pi
W_j^{-2}/2$, which is expected from the single-point-contact duality
as discussed below -- see Eqs.~(\ref{w1}) and (\ref{w2}). At large
voltages, the current saturates at the constant value
\begin{equation}
I = \frac{\pi \sigma_0}{2} {|\Gamma_1e^{i\kappa_V}+\Gamma_2|^2 \over
\Gamma_1 +\Gamma_2}\, . \label{4a1}
\end{equation}
At non-vanishing temperatures, the current $I$ depends linearly on
voltage $V$ at $V\ll T$. The corresponding linear conductance is
suppressed at large temperatures $T\gg \Gamma_{1,2}$ as
\begin{equation}
G = \frac{\pi \sigma_0 }{4T} {|\Gamma_1e^{i\kappa_V}+\Gamma_2|^2
\over \Gamma_1 +\Gamma_2}\, . \label{4a11}
\end{equation}

Behavior of the tunnel conductance $G\equiv I/V$ of the
interferometer at arbitrary temperatures is illustrated in Fig.~3,
which plots the conductance based on the Eqs.~(\ref{e25}) and
(\ref{4a10}) in the case of constructive interference, $\kappa_V=0$.
Note that for $\Delta t=0$ as assumed in this Section, the
conductance $G$ depends on the interference phase $\kappa_V$ only
through the amplitude $|\Gamma_1 e^{i\kappa_V} + \Gamma_2|$, which
gives the full 100\% modulation of $G$ for identical contacts, when
$\Gamma_1=\Gamma_2$, and suppression of the modulation with
increasing contact asymmetry. At $T\rightarrow 0$, the conductance
reflects the crossover from the regime of electron tunneling at
small voltages, characterized by $G\propto V^2$, to the regime of
quasiparticle tunneling at large voltages, where $G\propto 1/V$. The
conductance reaches maximum in the crossover region. The rate of
electron tunneling is enhanced by non-vanishing temperatures, so
that $G\propto T^2$, when $V\ll T\ll \Gamma_{1,2}$. At large
temperatures, $T> \Gamma_{1,2}$, the electron tunneling regime
effectively disappears, and conductance approaches the asymptotic
value (\ref{4a11}) that is independent of the voltage $V$ in the
range $V<T$.

\begin{figure}[htb]
\setlength{\unitlength}{1.0in}
\begin{picture}(3.,2.1)
\put(0.1,-0.15){\epsfxsize=2.8in \epsfbox{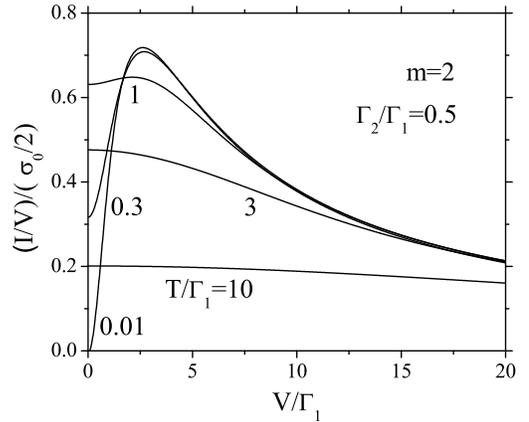}}
\end{picture}
\caption{Tunneling conductance of the symmetric
Mach-Zehnder interferometer formed between the edges
with filling factors $\nu_0=1/3$ and $\nu_1=1$ as a
function of the bias voltage $V$ between them for
several temperatures $T$ in the case of maximum
constructive interference, $\kappa_V=0$. The conductance
is calculated from \protect Eqs.~(\ref{e25}) and
(\ref{4a10}). The curves illustrate the low-$T$
crossover between the electron tunneling regime at low
voltages and tunneling of the edge-state quasiparticles
of charge $e/2$ at large voltages. The crossover is
manifested in the conductance peak at the intermediate
voltages which disappears with increasing temperature.}
\end{figure}

We note that the results for the current $I$ obtained above do not
depend on the average propagation time $\bar{t}$ between the
contacts, and are symmetric with respect to the interchange of the
contact indexes $1$ and $2$. It is therefore instructive to consider
the limit $x_1=x_2$, when Eqs.~(\ref{e22}) are transformed into
\begin{eqnarray}
\partial_t\xi_j(t)\!&=&\!2iv[w_j\psi^+(x_j,t)-w_j^*\psi(x_j,t)] \, ,
\nonumber \\ i\psi(x)|^{x_1+0}_{x_1-0}\!\!&=&\! \sum_j w_j \xi_j .
\label{4a2} \end{eqnarray}
Substitution into these equations of the operators $\psi(x,t)$ and
$\xi_{1,2}(t)$ in the following form
\begin{eqnarray}
\psi(x,t)&=&\int {d k \over 2 \pi} e^{ik(x-vt)}(\theta(x_1-x) \psi_k
\nonumber\\ &+&\theta(x-x_1) [S^{++}_k \psi_k +S^{-+}_k
\psi^+_{-k}]) \, , \nonumber \\ \xi_j(t)&=&\int {d k \over 2 \pi}
e^{ik(x-vt)}\xi_{j,k} , \label{4a3}
\end{eqnarray}
shows that
\begin{equation}
S^{-+}_k = {2ik\sum_j w_j^2\over (k+i\sum_j |w_j|^2)^2+|\sum_j
w_j^2|^2}\, . \label{4a4}
\end{equation}
Making use of this $S$-matrix element in the first part
of Eq.~(\ref{e24}), we find the expression for the
current which can be cast in the form similar to
Eq.~(\ref{e25}),
\begin{equation}
I={\Gamma_+ -\Gamma_- \over \Gamma_+ + \Gamma_-} [
I_{1/2}(V,\Gamma_-)-I_{1/2}(V,\Gamma_+) ]\, , \label{4a5}
\end{equation}
where $\Gamma_\pm=v(\sum_j|w_j|^2 \pm |\sum_j w_j^2|)$. In general,
$\Gamma_\pm$ differ from $\Gamma_{1,2}$, so that the two
current-voltage characteristics: for two different point contacts,
and for one ``combined'' contact, do not coincide. This shows that
although the current (\ref{e25}) is independent of the propagation
time $\bar{t}$, it does depend on the fact that $\bar{t}$ is
non-vanishing. Nevertheless, since $\Gamma_+ +\Gamma_-= \Gamma_1+
\Gamma_2$ and $\Gamma_+ -\Gamma_- =|\Gamma_1 e^{i\kappa_V} +
\Gamma_2|$, the two current-voltage characteristics have the same
large-voltage (\ref{4a1}) and large-temperature (\ref{4a11})
behavior. This large-energy equality is ``symmetric'' to the fact
that the low-energy asymptotics (\ref{e9}) of the electron tunnel
current for negligible $\Delta t$ is given by the single-contact
expression with the geometric sum of the electron tunneling
amplitudes.

\subsection{Bethe-ansatz solution}

The results for $\lambda=2$ discussed above can be
generalized to other values of $\lambda^2=2m$, for which
a thermodynamic Bethe-ansatz solution is known \cite{ba}
for a single-point tunneling contact. The solution
exploits a set of quasiparticle states describing local
$\vartheta_-(x)$ excitations and introduced through the
massless limit of the sine-Gordon model. These
quasiparticles are kinks, antikinks, and breathers of
the height defined by the sine-Gordon interaction and
equal to $\pi \lambda$. They remain interacting in the
massless limit as described by a bulk $S$ matrix, but
undergo separate one-by-one scattering at the point
contact described by a one-particle boundary $S$-matrix
\cite{za}. Their scattering at the two point contacts
occurs successively and independently at different
points, as follows from the dynamics of the local
fluctuations of the field $\vartheta_-(x)$ derived above
through application of the "unfolded" Dirichlet boundary
conditions. Therefore, the overall scattering is
described by the product of the two boundary $S$
matrices dependent on the phases $\kappa_1$ and
$\kappa_2$, respectively. To obtain these matrices from
the one found in Ref.~\cite{ba} in the case of
$\kappa=0$, we notice that each phase $\kappa_j$ in
Eq.~(\ref{e18}) results from the shift of $\vartheta_-$
by the constant $\kappa_j/\lambda$. Hence, the operators
$\exp(\pm i\lambda \vartheta_-/2)$ of the $\vartheta_-$
kinks and/or antikinks acquire just constant phase
factors $e^{\pm i\kappa_j /2}$. The boundary $S$-matrix
in Ref.~ \cite{ba} transforms then into
\begin{equation}
{\cal S}^{\pm \pm}_{j,k}={(a k/T_{jB})^{m-1}
e^{i\alpha_k}\over 1+i(a k/T_{jB})^{m-1} }\, , {\cal
S}^{- +}_{j,k} = {e^{i(\alpha_k -\kappa_{jV})}\over
1+i(a k/T_{jB})^{m-1} }\, , \label{e27}
\end{equation}
where the dimensional factor
\[ a=v {2\sqrt\pi\Gamma(1/[2(1-\nu)]) \over \nu
\Gamma(\nu/[2(1-\nu)])} \]
redefining the energy scales $T_{jB}$ is added into
Eq.~(\ref{e27}) to simplify the formulas below. The
tunneling current produced by the kink-antikink
transitions breaking the charge conservation takes the
following form for the two-point contact
\begin{equation}
I=\int_0^{\infty} v dk |(\hat{\cal S}_2\hat{\cal S}_1)^{-,+}|^2
n[f_+-f_-] \, .\label{e26}
\end{equation}
Notice that both the density of states $n(k,V)$ and the
distribution functions $f_\pm$ for kinks and antikinks,
are defined by the "bulk" of the system and do not
depend on the scattering at the point contacts. This
means that the tunneling current in Eq.~(\ref{e26})
takes the form that generalizes Eq.~(\ref{e25})
\begin{eqnarray}
{I \over V}={|T_{1B}^{m-1}e^{i\kappa_V}+T_{2B}^{m-1}|^2 \over
T_{1B}^{2(m-1)} -T_{2B}^{2(m-1)}} \times \;\;\;\;\;\;\;\;\;\;\;
\nonumber \\ \left[ G_{1/m}(V/T_{2B}, T/T_{2B}) -G_{1/m}(V/T_{1B},
T/T_{1B}) \right] , \label{e28}  \end{eqnarray}
where $G_{1/m}(V/T_{jB},T/T_{jB})$ is the universal
scaling function of the tunneling conductance of a
single-point contact between the two effective edges of
the filling factor $\nu=1/m$. This function has been
found \cite{ba} from the Bethe-ansatz solution, and at
zero temperature reduces to the low- and high-voltage
expansion series
\begin{eqnarray*}
G_{\nu}(s,0)=\sigma_0 \nu \sum_{n=1}^\infty c_n({1\over
\nu})
s^{2n({1\over \nu}-1)}\ \mbox{for}\  s<e^\Delta\, ,\\
G_{\nu}(s,0)=\sigma_0 \nu [1-\sum_{n=1}^\infty c_n(\nu)
s^{2n(\nu-1)}] \  \mbox{for}\  s>e^\Delta\, ,\\ c_n(\nu)
=(-1)^{n+1 }{\Gamma(\nu n+1)\Gamma(3/2)\over
\Gamma(n+1)\Gamma(3/2+(\nu-1)n)} ,
\end{eqnarray*}
where
\[ e^\Delta= (\sqrt{\nu})^{\nu/(1-\nu)}\sqrt{1-\nu}\, . \]
Substitution of these expansions into Eq.~(\ref{e28}) gives the
low-voltage asymptotics of the tunneling current for
$V<T_{jB}e^\Delta$ as
\[ {I \over V} = {\sigma_0 \over m } c_1(m) V^{2(m-1)}|
\sum_{j=1}^{2} T^{1-m}_{jB} e^{i\kappa_{jV}}|^2 ,
\]
and its large-voltage asymptotics for $V>T_{jB}e^\Delta$ as
\[ {I \over V} = {\sigma_0  \over m } c_1({1\over m})
V^{{2 \over m} -2}|\sum_{j=1}^{2} T^{m-1}_{jB}
e^{-i\kappa_{jV}}|^2{T_{1B}^{2-{2 \over m}}-T_{2B}^{2-{2 \over m}}
\over T^{2(m-1)}_{1B}-T_{2B}^{2(m-1)} } \, .
\]
The energy scales $T_{jB}$ are related to both correspondent
electron and quasiparticle tunneling amplitudes $U_j, W_j$ in the
same way
\begin{eqnarray}
T_{jB} & = & 2 D ({U_j\over\Gamma(1/\nu)}) ^{-{\nu\over 1- \nu}}
,\label{w1}\\
T_{jB} & = & {2 \over \nu} D ({W_j\over\Gamma(\nu)})^{{1\over 1-
\nu}}, \label{w2} \end{eqnarray}
as in the case of the individual point contact
\cite{weiss}. Substitution of Eq. (\ref{w1}) into the
low-voltage asymptotics reproduces exactly the
perturbative electron tunneling current Eq. (\ref{e9})
upon application of the identity $\sqrt \pi
\Gamma(2m)=2^{2m-1}\Gamma(m)\Gamma(m+1/2)$. On the other
hand, making use of (\ref{w2}) one can rewrite the
large-voltage asymptotics in terms of the quasiparticle
tunneling amplitudes
\[ {I \over V} = {|\sum_{j=1}^{2} W^{m}_{j}
e^{-i\kappa_{jV}}|^2 \over 2 \pi \Gamma(2/m) } {W_{1}^{2}-W_{2}^{2}
\over W^{2m}_{1}-W_{2}^{2m} } ({V \over m D})^{{2 \over m} -2}\, ,
\]
which agrees in the leading order with the calculation
in \cite{mz3}. Although the tunneling conductance
vanishes as a negative power of voltage (and
temperature), it always remains non-perturbative in the
quasiparticle tunneling amplitudes. This
non-perturbative dependence is a consequence of the
inherent resonance condition $|V\Delta t|<1$ in the
exact solution. Notice that the non-perturbative
behavior of the MZI takes place at large energies
contrary to the case of the antidot interferometer
\cite{b9}, where the resonant condition also leads to a
non-perturbative behavior, but at low energies. This
difference is related to formation of resonances around
the antidot, which can not be formed in the MZI, where
the two edges propagate in the same direction.

Several other general features follow directly from the
expression (\ref{e28}) for the current. Equation
(\ref{e28}) shows that the current interference has the
same dependence on both $V$ and $T$ as the function
$G_{1/m}$ of the single-point tunneling conductance.
This similarity holds only when $V\Delta t,T\Delta t\ll
1$. Indeed, as we have seen from Eq.~(\ref{e5}), in the
perturbative regime of electron tunneling, the condition
$V\Delta t\gg 1$ leads to the power-law suppression of
the interference current, and this suppression should
become exponential for $T\Delta t\gg 1$. Equation
(\ref{e28}) also shows that, similarly to
Eq.~(\ref{e9}), the visibility of the interference
pattern does not vary with temperature and voltage as
long as $V\Delta t,T\Delta t\ll 1$. In this regime, the
interference pattern produced by the dependence of the
current on the external magnetic flux has the same form
of the simple one-mode modulation, and is not affected
by the change from electron to quasiparticle tunneling.

To further clarify the typical patterns of the current modulation by
the interference phase $\kappa_V$, we consider Eq.~(\ref{e28}) in
the two limits: $T_{2B} \ll T_{1B}$ and $T_{2B} = T_{1B}$. In the
first case, expression for the tunneling conductance simplifies to
\begin{eqnarray}
G \simeq  \left[1+2\cos\kappa_V \left({T_{2B}\over
T_{1B}} \right)^{m-1}\right] \times \;\;\;\;\;\;\;
\nonumber \\ \left[ G_{1\over m} (V/T_{2B},
T/T_{2B})-G_{1 \over m} (V/T_{1B}, T/T_{1B})\right] ,
\label{w3}
\end{eqnarray}
which shows that for $T_{2B}<(T\  \mbox{or}\  V) < T_{1B}$, the
conductance exhibits weak oscillations of the amplitude
$U_1/U_2=(W_2/W_1)^m$ as a function of the external magnetic flux
close to the single-point-contact saturation value $\sigma_0/m$. For
$V,T<T_{2B}$, or if at least one of the energies is larger than
$T_{1B}$, the conductance goes to zero. Although the proportionality
of the amplitude of the interference oscillations to the $m$th
power, $W_2^m$, of the smaller quasiparticle tunneling amplitude has
entered Eq.~(\ref{w3}) throught the single-point-contact duality
relations (\ref{w1}) and (\ref{w2}) between $U_2$ and $W_2$, one can
also interpret it as a manifestation of the quasiparticle
statistics. Indeed, in the general quasiparticle tunneling model
described by the Lagrangian (\ref{e18}), the appearance of $W_2^m$
in the amplitude of the current oscillations is a mathematical
consequence of the Klein factor relations (\ref{e17}). In terms of
physics, it is also necessary in order to restore the $\Phi_0$
periodicity of the current in the external magnetic flux, since the
quasiparticle statistics implies that each tunneling of a
quasiparticle changes the effective flux for other quasiparticles by
$\Phi_0/m$. Therefore, in the MZI, the $W_2^m$ dependence of the
current oscillation amplitude, and the appearance of $e/m$
fractional charge in the quasiparticle shot noise discussed earlier,
both originate from the fractional statistics of the quasiparticles.

In the case of identical contacts $T_{2B} = T_{1B}$, and for $T=0$,
Eq.~(\ref{e28}) can be written as
\begin{equation}
G ={2 \cos^2(\kappa_V/2) \over m-1 }\ V\ \partial_V
G_{1/m}({V\over T_{B}}) \, . \label{w4}
\end{equation}
Its average over the magnetic flux oscillations is equal to the
oscillation amplitude, and also coincides \cite{banoise} with the
doubled shot noise $2\langle I^2\rangle (V,T_B)/V$ of the tunneling
current through the point contact divided by the voltage. At finite
temperature $T$, the linear conductance is also given by
Eq.~(\ref{w4}) with the voltage $V$ replaced by temperature $T$.

\begin{figure}[htb]
\setlength{\unitlength}{1.0in}
\begin{picture}(3.,2.3)
\put(0.0,-0.1){\epsfxsize=3.0in \epsfbox{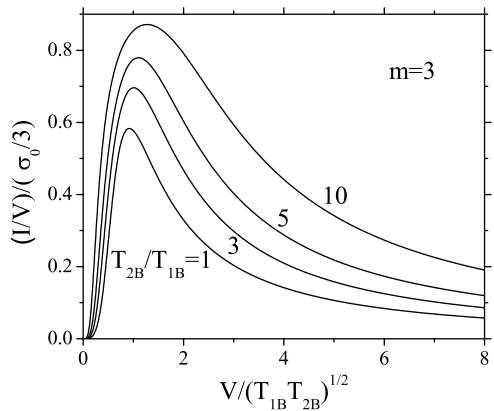}}
\end{picture}
\caption{Zero-temperature tunneling conductance \protect (\ref{e28})
of the symmetric Mach-Zehnder interferometer with $m=3$ (formed,
e.g., between the two edges with filling factors $\nu_0=1/3$) as a
function of the bias voltage $V$ for different degrees of the
asymmetry between the two interferometer contacts in the case of
maximum constructive interference, $\kappa_V=0$. The curves
illustrate the crossover between the electron tunneling at low
voltages and quasiparticles tunneling at large voltages. The
crossover region is seen as the conductance peak between the two
regimes. }
\end{figure}

The zero-temperature tunneling conductance at the
intermediate ratios $T_{2B}/T_{1B}$ calculated from
Eq.~(\ref{e28}) for $m=3$ is plotted in Fig.~4. The
conductance is shown in the case of maximum constructive
interference, $\kappa_V=0$. As discussed above, the
conductance depends on the interference phase $\kappa_V$
only through the prefactor $|T_{1B}^{m-1} e^{i\kappa_V}
+T_{2B}^{m-1}|$, so that the magnitude of the
interference current decreases monotonically with the
degree of asymmetry between the two contacts. Figure 4
shows that the width of the crossover region between the
electron tunneling at low voltages and quasiparticle
tunneling at large voltages increases with increasing
contact asymmetry. Simultaneously with the increasing
width of the conductance peak in the crossover region,
its height increases towards the conductance saturation
value $\sigma_0/m$. This behavior is consistent with the
simple qualitative picture of the total tunneling
current being the difference between currents in the two
point contacts. The larger the difference between the
two energy scales $T_{1B}$ and $T_{2B}$, the larger is
the voltage region where the conductance of the more
transparent contact already reached the saturation,
while the conductance of the less transparent contact
remains small.

\section{Summary and discussion}

Starting from the electron tunneling model of the
electronic Mach-Zehnder interferometer that is natural
at low voltages and/or temperatures, when the tunneling
is weak, we have calculated the quantum average of
current at all energies. The average current oscillates
as a function of magnetic flux with the period of one
flux quantum due to interference of tunneling electrons
at low energies and quasiparticles at high energies. The
low-energy calculation shows that the interference
oscillations are suppressed with increasing difference
$\Delta t$ of the propagation times along the two edges,
due to variations in the interference phases with energy
and/or momentum of the propagating excitations. The
tunneling current does not depend on the average
propagation time $\bar t$, and for $\Delta t=0$, in the
lowest order of the perturbation theory in the electron
tunneling amplitudes, does not distinguish the
geometrically different situations of $\bar t=0$ and
$\bar t \ne 0$.

Description of the strong tunneling regime which emerges
with increasing voltages and/or temperatures, has been
obtained by employing the instanton duality
transformation that introduces the quasiparticle
tunneling between the infinitely degenerate ground
states of the interferometer. The ground states are
defined by a choice of the branch of the phase produced
by interchange of electron tunneling processes at the
two point contacts. This phase gives the statistical
variation in the effective magnetic flux through the
interferometer. By minimizing the tunneling energy, we
have found the phase equal to $2 \pi m$ at $(t_0+t_1)D
\gg 1$, and derived the model
[Eqs.~(\ref{e17})-(\ref{e18})] of the quasiparticle
tunneling in the Mach-Zehnder interferometer for
arbitrary filling factors of the interferometer edges.
Although the tunneling terms at both contacts have
vanishing scaling dimension at high energies, the
perturbative treatment of the model at these energies is
possible only if $\Delta t V \gg 1$ or $\Delta t T\gg
1$, and the interference between the two tunneling
operators is suppressed. In this regime, the fractional
charge of the tunneling quasiparticles manifests itself
in the Schottky formula for the shot noise.

In the opposite limit of symmetric interferometer,
$\Delta t=0$, the model remains non-perturbative at high
energies, but allows the general exact solution which
describes the crossover from electron to quasiparticle
tunneling. The interference pattern of the current is
characterized by the single-harmonic modulation, which
is the same in both tunneling regimes, and is
independent of the voltage and temperature. The
modulation amplitude of the average current and also the
current shot noise carry signatures of the fractional
statistics of the quasiparticles.

It is interesting to compare our main exact result
(\ref{e28}) for the tunneling current with the solution
one would obtain by taking the zero-phase branch in the
interchange relations of the tunneling terms, which
follows from the minimization of energy at $(t_0+t_1)D
\ll 1$. In this situation, the geometry does not prevent
us from combining the two tunneling contacts together
into one effective point contact with the tunneling
amplitude $U_1+U_2e^{i\kappa_V }$. Its tunneling current
can then be found by substitution of this amplitude into
Eq.~(\ref{w1}) and the expressions for $G_{1/m}$. It has
the following low-energy expansion:
\[ {I\over V}={1 \over 2 \pi}\sum_{n=1}^\infty {c_n(m)\over m}({
|U_1+U_2e^{i\kappa_V }| \over \Gamma(m)})^{2n}({V\over 2
D})^{2n(m-1)}\ . \]
This expansion differs from the low-energy expansion of
current (\ref{e28}) already in the second lowest order
in the tunneling amplitudes. In this order, it involves
two oscillating harmonics as a function of $\kappa_V$.
This means that already the next order in the
perturbative expansion (discussed in Section II C) for
the electron tunneling current should distinguish the
geometry with $\bar t = 0$ from the geometry with $\bar
t \ne 0$. Even more noticeably, the difference in the
geometry of the tunneling contact would lead to the
different limiting values of the tunneling conductance
at large energies.

\begin{acknowledgments}

V.V.P. would like to thank Alvaro Ferraz for hospitality
and useful discussions during his stay at the
International Center for Condensed Matter Physics at the
University of Brasilia in Brazil, where a part of this
work was done. V.V.P. also acknowledges support of the
MCT of Brazil during this stay, of the ESF Science
Program INSTANS, and the Grant No. PTDC/FIS/64926/2006.

\end{acknowledgments}

\end{document}